# Crystal orientation-dependent oxidation of epitaxial TiN films with tunable plasmonics

Ruyi Zhang,[†,‡] Qian-Ying Ma,[†,§] Haigang Liu,[//] Tian-Yu Sun,[†] Jiachang Bi,[†] Yang Song,[†,‡] Shaoqin Peng,[†,‡] Lingyan Liang,[†] Junhua Gao, [*,†] Hongtao Cao,[†] Liang-Feng Huang,[*,†] and Yanwei Cao[*,†,‡]

[†] Ningbo Institute of Materials Technology and Engineering, Chinese Academy of Sciences, Ningbo 315201, China

[‡]Center of Materials Science and Optoelectronics Engineering, University of Chinese Academy of Sciences, Beijing 100049, China

[§]Nano Science and Technology Institute, University of Science and Technology of China, Suzhou, Jiangsu 215123, China

[//]Shanghai Synchrotron Radiation Facility (SSRF), Shanghai Advanced Research Institute, Chinese Academy of Sciences, Shanghai 201210, China

**ABSTRACT:** Titanium nitride (TiN) is a paradigm of refractory transition metal nitrides with great potential in vast applications. Generally, the plasmonic performance of TiN can be tuned by oxidation, which was thought to be only temperature-, oxygen partial pressure-, and time-dependent. Regarding the role of crystallographic orientation in the oxidation and resultant optical properties of TiN films, little is known thus far. Here we reveal that both the oxidation resistance behavior and the plasmonic performance of epitaxial TiN films follow the order of (001) < (110) < (111). The effects of crystallographic orientation on the lattice constants, optical properties, and oxidation levels of epitaxial TiN films have been systematically studied by combined high-resolution X-ray diffraction, spectroscopic ellipsometry, X-ray absorption spectroscopy, and X-ray photoemission spectroscopy. To further understand the role of crystallographic orientation in the initial oxidation process of TiN films, density-functional-theory calculations are carried out, indicating the energy cost of oxidation is (001) < (110) < (111), consistent with the experiments. The superior endurance of the (111) orientation against mild oxidation can largely alleviate the previously stringent technical requirements for the growth of TiN films with high plasmonic performance. The crystallographic orientation can also offer an effective controlling parameter to design TiN-based plasmonic devices with desired peculiarity, e.g., superior chemical stability against mild oxidation or large optical tunability upon oxidation.

Refractory plasmonic materials attract tremendous interest very recently due to their remarkable optical performance and durability in extreme environment.[1-3] As one paradigm of refractory plasmonic materials, titanium nitride (TiN) has been widely exploited in many applications including energy harvesting,[4-5] heat-assisted magnetic recording,[6] sensors,[7-8] photonic components,[9-13] photothermal therapy,[14] and medical imaging.[15] With the great advance in thin film synthesis technology, TiN films have been grown by sputtering,[16-18] pulsed laser deposition,[19-20] nitrogen-plasma-assisted molecular-beam epitaxy,[21] and atomic layer deposition.[22-23] One of the main challenges for TiN and many other transition metal nitrides is their oxidation during the thin film growth stage[24] and high-temperature operation (typically above 400 ºC) in plasmonic devices such as thermophotovoltaics and heat-assisted magnetic recording devices.[4, 25] Generally, slight oxidation can transform TiN into titanium oxynitride ($TiO_xN_y$) with red-shifted plasmon energy[26] and double epsilon-near-zero (ENZ) behavior, which can be used to enhance the nonlinear optical response.[24] However, over oxidation of metallic TiN can generate insulating titanium oxides (such as $TiO_2$)[27], degrading its plasmonic performance and leading to instability and even failure of optic components.[28]

To avoid oxidation, great efforts have been made by the scientific community. On one hand, an ultra-high vacuum condition is generally required during the growth of TiN thin films.[24] On the other hand, devices with TiN operated at high-temperature usually need to be carefully vacuumed,[25, 29] sealed in inert gases,[4] or covered by an optically transparent protective layer.[30] Previous studies have pointed out that the oxidation level of TiN is mainly controlled by temperature, oxygen partial pressure, and time.[31-33] Although several studies have indicated the impact of crystallographic orientation on the chemical reactivity of materials (metals, transition metal oxides, and organic materials),[34-37] its role in the oxidation of TiN thin films and resultant plasmonic performance is still unclear yet due to the lack of high-quality epitaxial TiN films with various controllable crystallographic orientations on the same substrates.

To reveal the role of crystallographic orientation in the oxidation and plasmonic performance of TiN films, we synthesized a series of high-quality (001)-, (110)-, (111)-oriented titanium (oxy)nitride epitaxial thin films using reactive magnetron sputtering. Unexpectedly, it is revealed that the order is (001) < (110) < (111) for both the oxidation resistance behavior and plasmonic performance in titanium (oxy)nitride thin films. Remarkably, (111)-oriented titanium (oxy)nitride thin films with bright golden color and excellent plasmonic properties can tolerate a record high oxygen partial pressure of $1\times10^{-5}$ Torr at 800 ºC during the deposition. Density-functional-theory (DFT) calculations are also carried out to understand the effect of crystallographic orientations on the initial oxidation process of TiN. Our results pave a way to fabricate oxidation resistant TiN films and enhance the high-temperature chemical stability of TiN optic components by controlling crystallographic orientation.

Titanium (oxy)nitride thin films were deposited on (001)-, (110)-, (111)-oriented $SrTiO_3$ (STO) substrates. It is noted that single-crystalline STO with cubic symmetry is a widely applied substrate for the epitaxy of perovskite oxide films but rarely used for the growth of titanium nitride thin films. Here, STO substrates were carefully selected because the large lattice mismatch (~ - 8 %) between titanium (oxy)nitride ($a$ ~ 4.2 Å) and STO ($a$ = 3.905 Å) favors a fast relaxation of titanium (oxy)nitride lattice, making us avoid considering the strain effect on substrates like MgO (a = 4.216 Å) having a small lattice mismatch with TiN film.[26] Another concern regarding the titanium (oxy)nitride films on MgO substrate is the strong overlap of film and substrate diffraction peaks, making it hard to determine the lattice parameter of titanium (oxy)nitride films. On the other hand, the study of TiN films on perovskite STO substrate is rare but important for integrating substantial transition metal nitrides with functional oxides for next-generation electronics. During the growth of pristine TiN films, high-pure $N_2$ gas (0.02 Torr with 5N purity) was utilized as the reactive gas with base vacuum pressure pumped at ~ $3\times10^{-8}$ Torr. To synthesize oxidized TiN films, partial pressure of ~$1 \times 10^{-5}$ Torr pure oxygen gas was added into the deposition chamber during the growth. In this work, for clarity, the pristine and oxidized (001)-oriented TiN films are labeled as 001 and 001*, respectively, which are similar for the (110)- and (111)-oriented films. It is noted that all growth parameters (substrate temperature, RF power, $N_2$ pressure, work distance, deposition time, *etc.*) were fixed except that the oxidation conditions varied for different batches of films. Moreover, to precisely study the behavior of orientation-dependent oxidation, (001)-, (110)-, and (111)-oriented STO substrates were mounted to the sample stage at one time, and thus the corresponding titanium (oxy)nitride films were synthesized at the same time. The thickness of all thin films is around 120 nm.

## RESULTS AND DISCUSSION

As schematically shown in Figure 1a, the surface termination of ideal TiN strongly depends on the crystallographic orientation. For example, the surface configuration of (001)-oriented TiN consists of cross-linked Ti and N atoms, whereas straightly aligned Ti-N bonding can be seen at the surface of (110)-oriented TiN. Interestingly, for the (111)-oriented TiN, there are two different surface configurations, i.e., Ti-terminated (orange plane) and N-terminated (blue plane). We find that slight oxidation during the thin film growth stage will not dramatically affect crystal symmetry and growth behavior of titanium (oxy)nitride thin films, because O atoms can be easily doped into the rock-salt TiN matrix as the anion solutes due to the very close lattice constants between TiO ($a$ = 4.177 Å) and TiN ($a$ = 4.244 Å). [38-40] As seen in Figure S1, diffraction peaks of all titanium (oxy)nitride thin films in the full-range $2\theta$-$\omega$ scans remain the same crystal indices as STO substrates, indicating preferential orientation growth for thin films without the trace of any secondary phase (e.g., anatase or rutile $TiO_2$) in the diffraction patterns. However, all diffraction peaks of the oxidized TiN films shift towards high diffraction angles with respect to the pristine cases (Figure 1b and Figure S1), signifying the reduction in lattice constants. Due to the slightly smaller lattice constant of TiO (4.177 Å) compared to that of TiN (4.244 Å), the partial replacement of $N^{3-}$ anions by $O^{2-}$ anions in TiN can result in the slightly smaller lattice constant of $TiO_xN_y$ samples.[41] The most prominent lattice constant reduction occurs in (001)-oriented films, which corresponds to a distinct diffraction angle upshift of ~ 0.3° after oxidation (Figure 1b). In comparison, (110)- and (111)-oriented oxidized TiN films experience less and the least lattice constant reductions, respectively, among the three kinds of films.

The epitaxial growth and lattice constants of titanium (oxy)nitride thin films can be characterized by the reciprocal space mappings (RSMs). Phi scans were also performed to further verify the cube-on-cube epitaxial growth for TiN thin films on STO substrates (Figure S2). Figure 1c shows the RSMs of the oxidized TiN thin films with (001), (110), and (111) orientations measured around (113), (222), and (113) diffractions, respectively. The mosaic dispersion of the film RSM peaks (lower left corner) confirms the expected lattice relaxation due to large lattice mismatch. Based on the RSM (113) data of (001)-oriented film, the (110) and (001) layer spacings can be determined to be 2.981 Å and 4.216 Å, respectively. In the same way, the (001) and (110) layer spacings are 4.229 Å and 2.990 Å for the (110)-oriented film, and the (11-2) and (111) layer spacings are 1.730 Å and 2.447 Å for the (111)-oriented films, respectively. According to these layer spacings, as expected, we find the lattice of all films to be cubic without any detectable sign of epitaxial strain, and the lattice constants of the (001)-, (110)-, and (111)-oriented oxidized TiN films are 4.216 Å, 4.228 Å, and 4.238 Å, respectively, which are apparently different under the same oxidation conditions. In comparison, according to RSMs shown in Figure S3, the lattice constants of the (001)-, (110)-, and (111)-oriented pristine TiN films are 4.243 Å, 4.244 Å, and 4.244 Å, which are nearly the same. The details about RSM analyses on all titanium (oxy)nitride thin films can be referred to Table S1.

It has been proven that Vergard's law was successful to describe the linear relationship between the lattice constant of a solid solution and the relative content of its constituents.[42] Then the oxygen content can be estimated by following Vergard's law and assuming that the cationic or anionic vacancies are not predominant ($x + y = 1$).[42] As shown in Figure 1d, the pristine TiN films all hold lattice constants very close to the bulk one (4.244 Å), suggesting the consistent chemical stoichiometry and negligible account of oxygen. In contrast, the oxygen content in the (001)-, (110)-, and (111)-oriented oxidized TiN films is 42%, 23%, and 9%, respectively.

Next, we investigate the crystallographic orientation and oxidation dependent optical properties of all titanium (oxy)nitride films by comparing their optical image colors, reflectivities, and dielectric functions (Figure 2a-c). As seen by the optical images in Figure 2a, there are visible color differences among the titanium (oxy)nitride epitaxial films, especially for the oxidized cases. The pristine TiN films show typical golden colors with slight differences in brightness, whereas oxidation darkens the golden color of 110* and 111* TiN films and even turns the color of the 001* film into dark orange, which seriously deviates from the golden color of plasmonic TiN in literature.[20-21] The different colors among the films (Figure. 2a) also closely correlate with their reflection spectra as shown in Figure 2b. The reflectivities ($R$) are calculated by using the formula $R = \frac{(n-1)^2+k^2}{(n+1)^2+k^2}$,[21] where $n$ and $k$ are the real and imaginary parts of the complex refractive index, respectively, derived from spectroscopic ellipsometry (SE) measurements. As seen, the 001* film possesses a constantly strong optical absorption (low $R$) across a wide wavelength range. However, all the other films only exhibit strong optical absorption of the visible lights and high reflectivity in the near-infrared-wavelength range (i.e. reflectivity > 80% above 1000 nm), which are the typical features for the plasmonic TiN.[21] Reflectivity increases in the order of (001) < (110) < (111), which is consistent with the brightness showed in Figure. 2a. The highest reflectivity reaches ~ 94% on the 111 film. Any oxidation tends to lower the reflectivity, which is most prominent on the 001* film.

The corresponding dielectric functions of all films were also compared with each other (Figure 2c,d). Drude-Lorentz model, a well-established model to simulate the contributions from the intraband and interband transitions in plasmonic TiN, is used to fit the obtained SE data.[20-21, 26] Since all films are thick ($t$ = 120 nm) and opaque, a two-layer structure consisting of air and TiN is used to extract the dielectric functions.[21] The fitting details can be referred to Figure S4 in Supporting Information. As seen in Figure 2c, the real part ε′ of permittivity in titanium (oxy)nitride thin films crosses from positive to negative at wavelengths $\lambda_p$ (corresponding to the screened plasma frequency $\omega_p$), which are red-shifted upon oxidation regardless of crystallographic orientation. The $\lambda_p$ are 470, 476, 477, 497, 500, and 587 nm for the 111, 111*, 110, 110*, 001, and 001* films, respectively, corresponding to $\hbar\omega_p$ of 2.64, 2.61, 2.60, 2.50, 2.48, and 2.12 eV. Except for the 001* film, the $\hbar\omega_p$ of all other films are close to the value of 2.65 eV for the plasmonic TiN.[43] Oxidation also leads to less negative ε′ and more positive ε″ (imaginary part of permittivity) in the near-infrared wavelengths for the (110)- and (111)-oriented films (Figure 2c,d), suggesting decreasing metallicity and increasing optical loss with oxidation. With the thin film growth parameters fixed, ε′ are the most negative in (111)-oriented films and the least negative in (001)-oriented films. The highest metallicity and relatively low optical loss are found in the 111 film, which is then followed by the 111* films. Their plasmonic performance is comparable to that of high-quality TiN films in literature.[20-21] Despite having the least negative in ε′, the 001* film shows a broadband ENZ behavior (|ε′| < 5 at 300 -1600 nm), which is very appealing in nonlinear optical applications.[44] Such optical behavior is similar to the dielectric properties of titanium oxynitride reported previously.[24, 45] The optical properties of thin films are closely connected with their electrical parameters. Therefore, we also measured their electrical parameters using the standard Van der Pauw 4-probe method through a Hall effect measurement system. As shown in Figure S5 (Supporting Information), the conductivity, carrier concentration, and carrier mobility all follow the same sequence: (001) < (110) < (111) for the pristine TiN films. The anisotropic electrical and optical properties are mainly attributed to different crystalline qualities as evidenced by the rocking curves shown in Figure S6. The room temperature carrier densities ($3.4 - 5.2 \times 10^{22}$ $cm^{-3}$) of the pristine TiN films are at the same level as that ($\sim 10^{22}$ $cm^{-3}$) of the typical plasmonic TiN.[20-21] Oxidation tends to lower both carrier concentration and mobility simultaneously. The results from Hall effect measurements agree very well with the trends shown by SE measurements.

The plasmonic performances of all titanium (oxy)nitride films in Figure 2 are further evaluated by the figure of merit for surface plasmon polariton ($FOM_{SPP}$) defined as $(\varepsilon')^2/\varepsilon''$. As seen in Figure 3, except for the 001* film with ENZ behavior, all other films show the values of $FOM_{SPP}$ larger than 1 at wavelengths > 600 nm, indicating the promising potential in SPP waveguide applications. Above wavelengths of 500 nm, $FOM_{SPP}$ is always the highest in (111)-oriented films while the lowest in (001)-oriented films with the same growth conditions, and oxidation tends to weaken $FOM_{SPP}$ for all orientations. Combining the analysis results from the SE and electrical measurements, we can conclude that the plasmonic performance and metallicity of titanium (oxy)nitride films follow the same order of (001) < (110) < (111) for both the pristine and oxidized films. The underlying reason for the inherited high plasmonic performance and metallicity of the oxidized 111* film is due to its superior oxidation resistance, which will be further demonstrated later. The oxidation conditions ($P_{O2} = 1 \times 10^{-5}$ Torr at 800 ºC) for the plasmonic 111* film are much harsher than the conventional conditions ($P_{O2} < 2 \times 10^{-8}$ Torr at 600 ºC) used to prepare $TiO_xN_y$ with double ENZ behavior.[24] Therefore, the crystallographic orientation (111) is ideal for the fabrication of TiN based plasmonic devices with superior chemical stabilities against mild oxidation. Furthermore, an endurance against the exposure to $P_{O2} = 1 \times 10^{-5}$ Torr at 800 ºC for the (111)-oriented titanium (oxy)nitride epitaxial thin films with high plasmonic performance indicates that the very low base pressure (e.g., $\sim 10^{-8}$ Torr) is not necessary in this case, which can efficiently lower the technical threshold to prepare high-quality titanium (oxy)nitride thin films. Since the (001)-oriented TiN is more susceptible to oxidation, thus it is most suitable as tunable optical materials for nonlinear optics and multi-resonant plasmonic applications.[45]

To further explore the orientation-dependent oxidation behaviors, oxidation levels and electronic structures of all titanium (oxy)nitride thin films are investigated via the element-specific X-ray absorption spectroscopy (XAS) and X-ray photoemission spectroscopy (XPS). XAS with surface-sensitive total-electron-yield (TEY) mode is a very powerful technique to probe the TiN surface, which quickly forms a few nanometers-thick native oxidation layer after exposure to the atmosphere.[46-47] As shown in Figure 4a, the O K-edge XAS spectra arise from electron transition from O 1*s* to the unoccupied O 2*p* states, which also carry Ti band characters via hybridization.[47] The predominant features *A, B, C, D* resemble those in the titanium oxynitride films prepared with $P_{O2} < 5 \times 10^{-5}$ Torr in the literature, demonstrating the hybridized O 2*p*- Ti 3*d* (*A* and *B*) and O 2*p*- Ti 4*sp* states (*C* and *D*).[47-49] The oxidation levels at the pristine TiN film surfaces increase in the order of (001) < (110) < (111) as evidenced by their XAS intensities, indicating that the (111) and (001) surfaces are the most resistive and vulnerable to natural oxidation, respectively. The similarly low XAS intensities for both 111 and 111* films suggest superior oxidation resistance and chemical stability of (111) surface against mild oxidation. However, (001) and (110) surfaces are comparatively less stable, and escalated oxidations tends to occur at these surfaces, resulting in higher XAS intensities in the 001* and 110* TiN films. We notice that the peak energies for feature *A* and *B* are almost invariable for all films, indicating relatively stable O 2*p*- Ti 3*d* interactions. The energy shifts for features *C* and *D* are faint among the oxidized TiN films but very distinct among the pristine TiN films. Generally, the dispersive features *C* and *D* are more sensitive to the long-range order, which can cause energy shift by enhancing the Ti-O interactions.[47, 49] The obvious energy shifts for features *C* and *D* suggest the enhanced Ti-O interactions might benefit from the enhanced long-range order via oxidation in the film surfaces.

To further unveil the oxidation levels beneath the highly oxidized surface layer, the Ti 2*p* XPS spectra are recorded after etching. For the TiN surfaces with native oxidation layer, Ti 2*p* spectra generally consist of three pairs of $2p_{3/2}$ and $2p_{1/2}$ spin-orbital components, which can be assigned to Ti-O (458.3 eV, 464.0 eV), O-Ti-N (456.8 eV, 462.5 eV), and Ti-N (454.8 eV, 460.6 eV) bondings, respectively.[50] But for the clean TiN surface without oxidation, broad shake-ups with a full width at half maximum (FWHM) of ~ 3.8 eV appear at the shoulders of peaks assigned to Ti-N bonding (455 eV, 460.9 eV) because of that Ti 3*d*-N 2*p*-hybridized states near the Fermi level could relax selection rules and allow a multitude of transitions to different states.[50-51] Here, we find the Ti 2*p* spectra of the pristine TiN films in Figure 4b can be well deconvoluted into one doublet (454.9 eV for $2p_{3/2}$ and 460.8 eV for $2p_{1/2}$) assigned to Ti-N bonding and two broad shake-ups (FWHM ~ 4 eV), consistent with the spectra of clean TiN.[50-51] The spin-orbit coupling of 5.9 eV also agrees very well with that (~ 5.9 eV) of TiN in the literature.[50-51] Similar to the pristine TiN films, the XPS spectra of the oxidized films in Figure 4c also possess features interpreted as Ti-N bonding and its shake-ups. However, it should be noticed that the relative intensity rises in the regions (455.3 - 457.7 eV and 460.9 - 463.3 eV) between Ti-N peaks and corresponding shake-up peaks as the orientation change from (111) to (001). These regions can only be assigned to O-Ti-N bonding.[26] Therefore, the oxidation levels beneath the surface layer for the oxidized TiN films also follow the order of (111) < (110) < (001), which matches very well with the trend shown in Figure 1d. The XPS O 1s spectra of oxidized TiN films shown in Figure S7 further demonstrate that the oxygen content follows the order of (111) < (110) < (001).

Finally, we performed DFT calculations to understand the microscopic mechanism for crystallographic orientation dependent oxidation of TiN films. The oxidation of TiN can be viewed as a process of continuous oxidation of TiN surfaces while the thin films keep growing. The initial oxidation process of the TiN surface mainly involves two critical steps, i.e., (1) the adsorption of O atoms on the TiN surface, and (2) the exchange of subsurface N by O adatom.[52]

We consider both configurations with two adsorbed O atoms (molecule or dissociated adatoms, denoted as $O_{2\text{-ad}}$) and one O adatom plus another O atom exchanged with a subsurface N atom (denoted as $O_N$).[52] The adsorption stability/tendency of a free $O_2$ molecule approaching a TiN surface can be described by its adsorption energy ($E_{ad}$).[53-54] The tendency of an O adatom to enter into the TiN surface can be indicated by the energy cost ($\Delta E$) for exchanging an O adatom with a subsurface N atom, as expressed by $\Delta E = E_{ON} - E_{ad}$, where $E_{ON}$ is the electronic energy after the O-N exchange.[52] Many possible adsorption configurations for each case are provided for the initial structural screening, and the most stable one is located after DFT relaxations (Figure S9~S11, Supporting Information). In addition, the (111) orientation has two possible surface terminations, i.e., the $(111)_N$ surface with N-termination and $(111)_{Ti}$ surface with Ti-termination, both of which are also considered in our calculations.

Figure 5a-c show the most stable atomic structures in the oxidation steps (1) and (2), as well as their corresponding $E_{ad}$ and $\Delta E$, for different crystallographic orientations. As seen, $E_{ad}$ has a remarkable dependence on the surface type, with a trend of $(111)_{Ti} < (110) < (001) < (111)_N$. Generally, more unpaired surface electrons that are created by breaking the bulk lattice bonds lead to a more reactive surface, and then higher stability of O adatom (lower $E_{ad}$) on its surface.[55] The calculated surface energies ($\gamma$) follow the trend (001) < (110) < (111) (Figure S8, Supporting Information), suggesting the Ti sites on (001) and (111) surfaces being the least and most reactive, respectively. This mechanism explains the trend of $(001) > (110) > (111)_{Ti}$ in $E_{ad}$ (Figure 5a-c). Our DFT calculations also indicate that the O-O bonding tends to be broken by the strong interaction from the surface Ti atoms, making the dissociation of an $O_2$ molecule occur spontaneously with very low adsorption energies (Figure S9~S11). All of the three TiN surfaces, except for $(111)_N$, have large negative $E_{ad}$ values ranging from -3 to -12 eV, indicating that the surface Ti atoms will be readily covered by O adatoms in oxidizing gaseous environments. The front N anions on the N-terminated $(111)_N$ surface have a quite low reactivity in terms of $O_2$ adsorption, and the $E_{ad}$ value is as small as -0.09 eV. Thus, the inevitable exposure of $(111)_N$ surface during a preparation or oxidation process (e.g., the alternating oxidation of the Ti and N surface layers) tends to largely inhibit the oxygen adsorption behavior.

The continuous oxidation and formation of rocksalt titanium oxynitride mean oxygen atoms have to be firmly bonded to Ti atom in the lattice, which requires the exchange of subsurface N by O adatom. As shown by the calculated $\Delta E$ value for each orientation (Figure 5a-c), the O-N exchange process is only exothermic (i.e., $\Delta E < 0$) on the (001) surface, but endothermic on the (110) and (111) surfaces. The trend in $\Delta E$ is (001) < (110) < (111) because it is more difficult to exchange a more stable O adatom with a lattice N atom. The O-N exchange tendency can well imply the proceeding tendency of the surface oxidation, and the calculated results are consistent with the experimental observations in this work that considerable oxidation is only observable on (001) orientation and (111)-oriented TiN films show the best oxidation-resistant behavior.

## CONCLUSION

To summarize, we have synthesized a series of high-quality titanium (oxy)nitride epitaxial films by using reactive RF sputtering technique and systematically studied the effect of crystallographic orientation on the lattice constant, optical property, oxidation level of these films. It was revealed that both oxidation resistance behavior and plasmonic performance of titanium nitride films follow the order of (001) < (110) < (111). Remarkably, (111)-oriented titanium (oxy)nitride films with high plasmonic performance can endure a record high $O_2$ partial pressure of $10^{-5}$ Torr at 800 ºC during the film growth process, which provides a technical advantage in growing high-quality plasmonic TiN with much lower requirements for the vacuum system. Through DFT calculations, we revealed that the high energy cost for the O-N exchange process on (111)-oriented TiN surface endows the superior oxidation resistance. The understanding gained from this work can pave a flexible route towards designing optical devices based on plasmonic TiN thin film with desired peculiarity, either better chemical stability against mild oxidation or larger optical tunability upon oxidation.

## METHODS

**Thin Film Growth.** A home-made RF magnetron sputtering system was used to synthesize titanium (oxy)nitride thin films on (001)-, (110)-, and (111)-oriented $SrTiO_3$ substrates. Initially, the base pressures were pumped at $3\times10^{-8}$ Torr. After that, the pipelines and deposition chamber were purged with nitrogen (99.999% purity) gas and then pump down to ~ $3\times10^{-8}$ Torr again. These procedures were done to decrease residual oxygen in the chamber before subsequent depositions. The pristine TiN films were deposited at 800 °C under pure nitrogen pressure of 0.02 Torr with a flow rate of 3.2 sccm. All samples were mounted to a rotatable SiC absorber heated by a laser. The temperatures were monitored at the back side of SiC absorber by an infrared pyrometer. The rotation speed for the sample holder was kept at 5 rpm during deposition. The output power of laser module was kept at 26.3 % of full power (200 W) to maintain a constant temperature of 800 ℃. The RF generator power was kept at 100 W during deposition. The work distance between sputter gun and substrate is 20 cm. Additional $O_2$ gas (99.999% purity) with partial pressure of $1\times 10^{-5}$ Torr was introduced into the chamber through a fine adjustment valve to prepare oxidized TiN thin films.

**Structural and Chemical Characterization.** The $2\theta$-$\omega$ scans and reciprocal space mappings were performed on a high-resolution X-ray diffractometer (Bruker D8 Discovery) using monochromatic Cu $K_{\alpha1}$ radiation with a wavelength of 1.5406 Å. The film thicknesses were determined by X-ray reflectivity. The surface electronic structures and oxidation levels of all films were investigated by X-ray absorption spectroscopy with total-electron-yield detection mode at BL08U1A beamline in Shanghai synchrotron radiation facility (SSRF) after all films were exposed in air for about 2 months. The chemical compositions beneath the surface layer were detected by X-ray photoemission spectroscopy (Kratos AXIS Supra) at acceptance angles of 45° after all film surfaces were etched by Ar ions.

**Optical and Electrical Properties Characterization.** The spectroscopic ellipsometry measurements were performed at a variable-angle spectroscopic ellipsometer (J. A. Woollam M-2000DI) with wavelengths ranging from 400 nm to 1600 nm. The electrical transport of films was measured by using a Hall effect measurement system using the standard Van der Pauw 4 probe method from 300 K to 400 K.

**Density-Functional-Theory Calculations.** The density-functional theory is implemented in the Quantum ESPRESSO code package.[56] The spin-polarized Perdew-Burke-Ernzerhof functional in the generalized gradient approximation is used to describe the electronic exchange and correlation,[57-58] and the projector augmented wave method is used to express the electronic wave functions and potentials.[59] The convergence thresholds for the atomic force and electronic energy are 0.01 eV/Å and $2\times10^{-8}$ Ry, respectively. According to our systematic calculations (Figure S8, Supporting Information), periodic structural slabs of 5, 5, 6 atomic layers are thick enough to model the (001), (110), and (111) surfaces in the DFT calculations here, and the neighboring slab images are separated by a vacuum spacing of ~20 Å that can effectively exclude any inter-slab interaction. The Brillouin zones of the periodic (001), (110), and (111) slabs are sampled by the reciprocal Monkhorst−Pack grids of 6×6×1, 6×4×1, and 5×4×1, respectively.[60] The intrinsic stability of a TiN surface can be well indicated by its surface energy (γ for each unit area), as expressed as $\gamma = \frac{1}{2A}(E_{slab} - N \times E_{bulk})$,[61] where $E_{\mathrm{slab}}$ and $E_{\mathrm{bulk}}$ are the electronic energies of the slab structural model (with 2 surfaces) and the bulk unit cell, respectively; $A$ is the area of each surface unit, and the prefactor of 2 originates from the two surfaces in the slab structure; $N$ is the number of the unit cell across the slab and can also indicate the thickness. The adsorption stability/tendency of a free $O_2$ molecule approaching a TiN surface can be well described by its adsorption energy ($E_{\mathrm{ad}}$), which is expressed as $E_{ad} =$ $E_{tot} - E_{slab} - E_{O2}$, where $E_{tot}$, $E_{slab}$, and $E_{O2}$ are electronic energies of the systems of a slab with two O adatoms, a clean slab, and a free $O_2$ molecule, respectively. A more negative $E_{ad}$ indicates a more favorable adsorption. After the adsorption of O atoms, the initial oxidation process of the surface usually starts with the entry of an O adatom into the TiN surface, and the tendency can be indicated by the energy cost ($\Delta E$) for exchanging an O adatom with a surface N atom.

## ASSOCIATED CONTENT

### Supporting Information

Supporting Information is available online.
Full-range $2\theta$-$\omega$ scans, phi scans, detailed RSM analyses, rocking curves and temperature-dependent electrical transport of titanium (oxy)nitride films, fitting details of SE data, XPS O1*s* spectra, surface structures and energies in the initial oxidation process of TiN calculated by DFT.

## AUTHOR INFORMATION

### Corresponding Author

*E-mail: gaojunhua@nimte.ac.cn, huangliangfeng@nimte.ac.cn, ywcao@nimte.ac.cn

### Author Contributions

R. Zhang and Q. Ma contributed equally to this work. R. Zhang and Y. Cao conceived the project. R. Zhang, J. Bi, S. Peng, and Y. Song prepared the samples and performed the HRXRD, SE, Hall, and XPS measurements. R. Zhang, Y. Song, and H. Liu carried out XAS measurements and analyses. Q. Ma, T. Sun, and L. Huang performed the DFT calculations and analyzed the data. J. Gao, L. Liang, and H. Cao discussed the fittings and analyses of SE data. All authors discussed the data and contributed to the manuscript.

### Notes

The authors declare no competing financial interest.

## ACKNOWLEDGMENTS

We acknowledge the XAS experiments support from BL08U1A of Shanghai Synchrotron Radiation Facility. This work was supported by the National Natural Science Foundation of China (Grant No. 11874058 and U2032126), the Pioneer Hundred Talents Program of Chinese Academy of Sciences, the Ningbo 3315 Innovation Team, and the Ningbo Science and Technology Bureau (Grant No. 2018B10060). This work was partially supported by Youth Program of National Natural Science Foundation of China (Grant No. 12004399), China

Postdoctoral Science Foundation (Grant No. 2018M642500), and Postdoctoral Science Foundation of Zhejiang Province (Grant No. zj20180048).

**REFEERENCES**

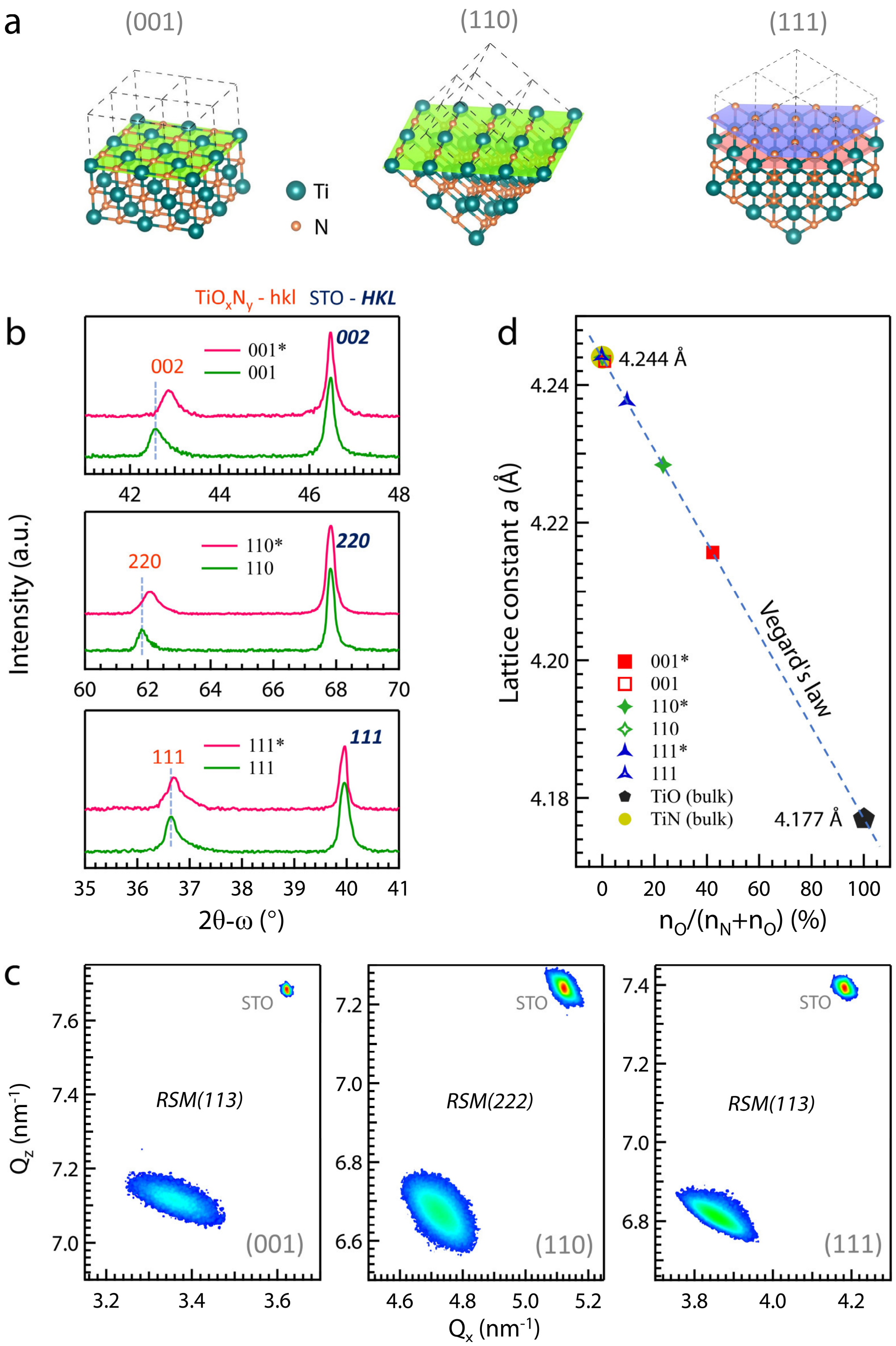

**Figure 1.** (a) Schematic of (001)-, (110)-, and (111)-oriented surfaces of TiN. It is noted that (111)-TiN either have Ti-terminated (orange plane) or N-terminated (blue plane) surfaces. (b) *2θ-ω* data of pristine and oxidized TiN films along different orientations. (c) RSM patterns around (113), (222), and (113) diffractions for (001)-, (110)-, and (111)- oriented $TiO_xN_y$ films, respectively. The patterns of $TiO_xN_y$ films are at the left bottom, whereas the features of STO substrates are at the right top. (d) Calculated lattice constants and estimated oxygen contents ($n_O/(n_N+n_O)$) according to Vegard's law. The oxygen content of 0 % and 100 % corresponds to pure TiN and TiO, respectively.

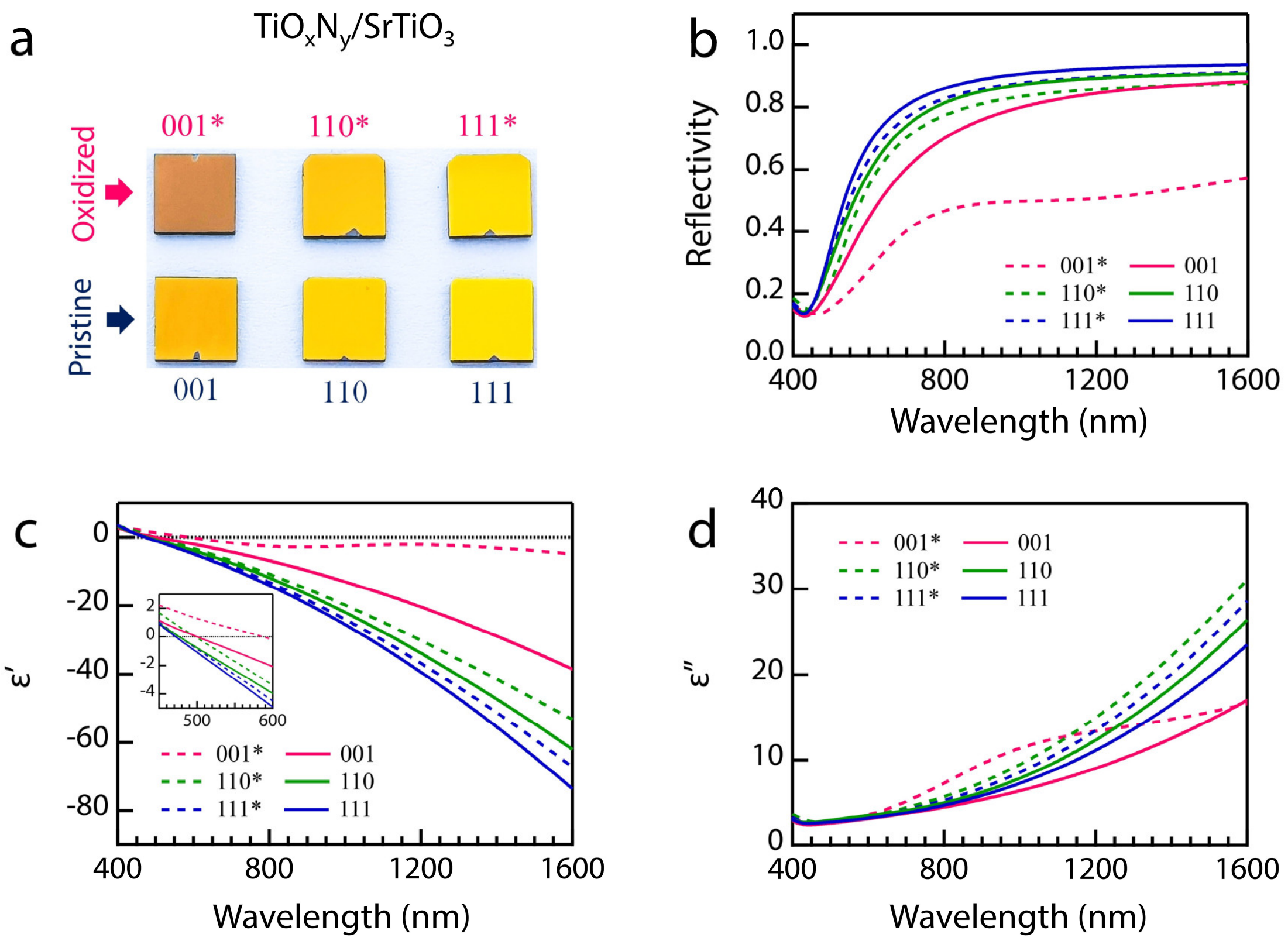

**Figure 2.** (a) Optical image of the pristine and oxidized TiN films with (001), (110), and (111) orientations. (b) Derived reflectivity of all titanium (oxy)nitride thin films. (c, d) Extracted dielectric functions (real part ε′ and imaginary part ε″) from SE measurements for all films. The inset in (c) is an enlarged view around the crossover wavelength $\lambda_P$.

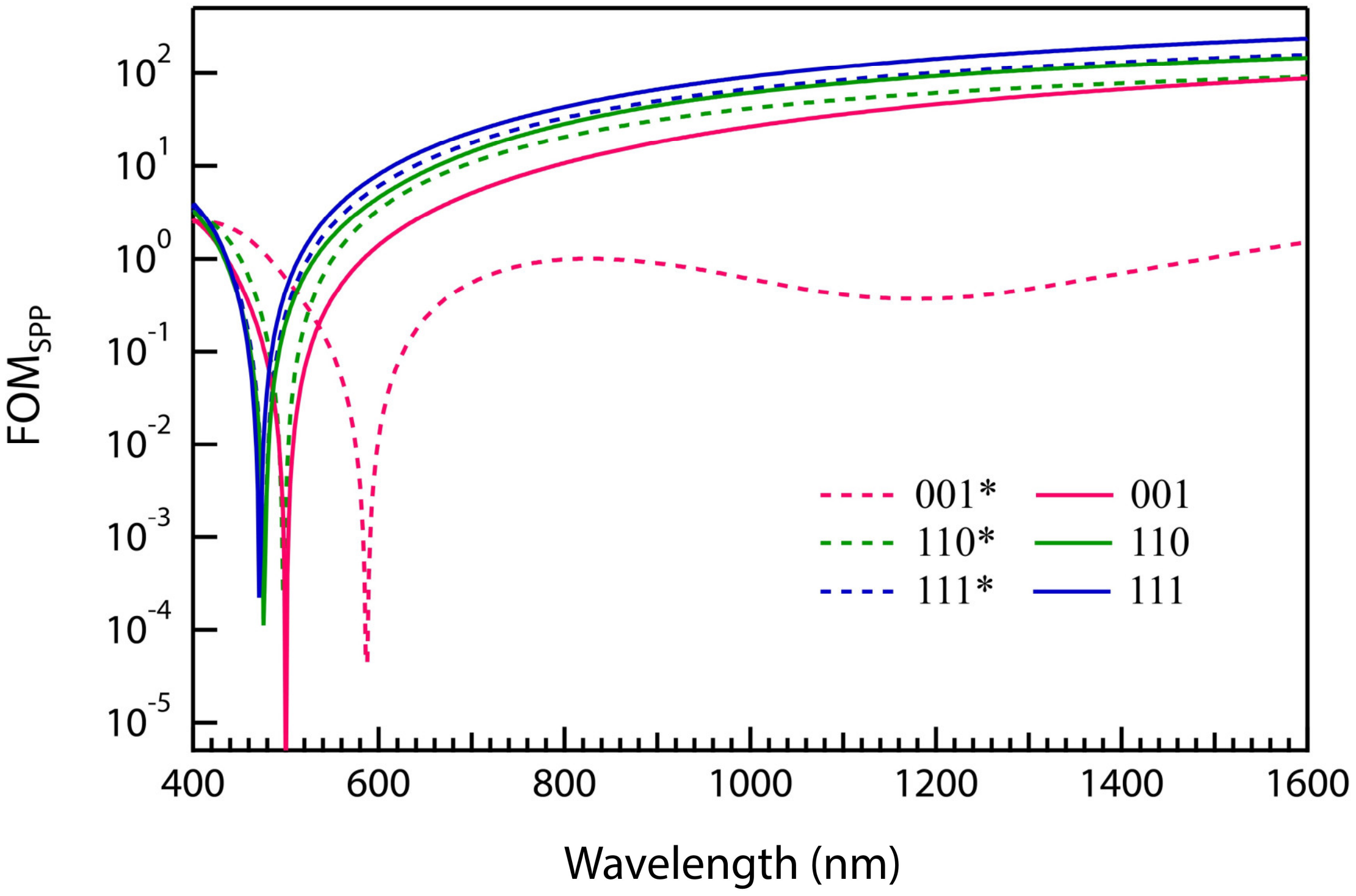

**Figure 3.** Figure of merit for surface plasmon polariton ($FOM_{SPP} = (\varepsilon')^2/\varepsilon''$) of pristine and oxidized TiN films.

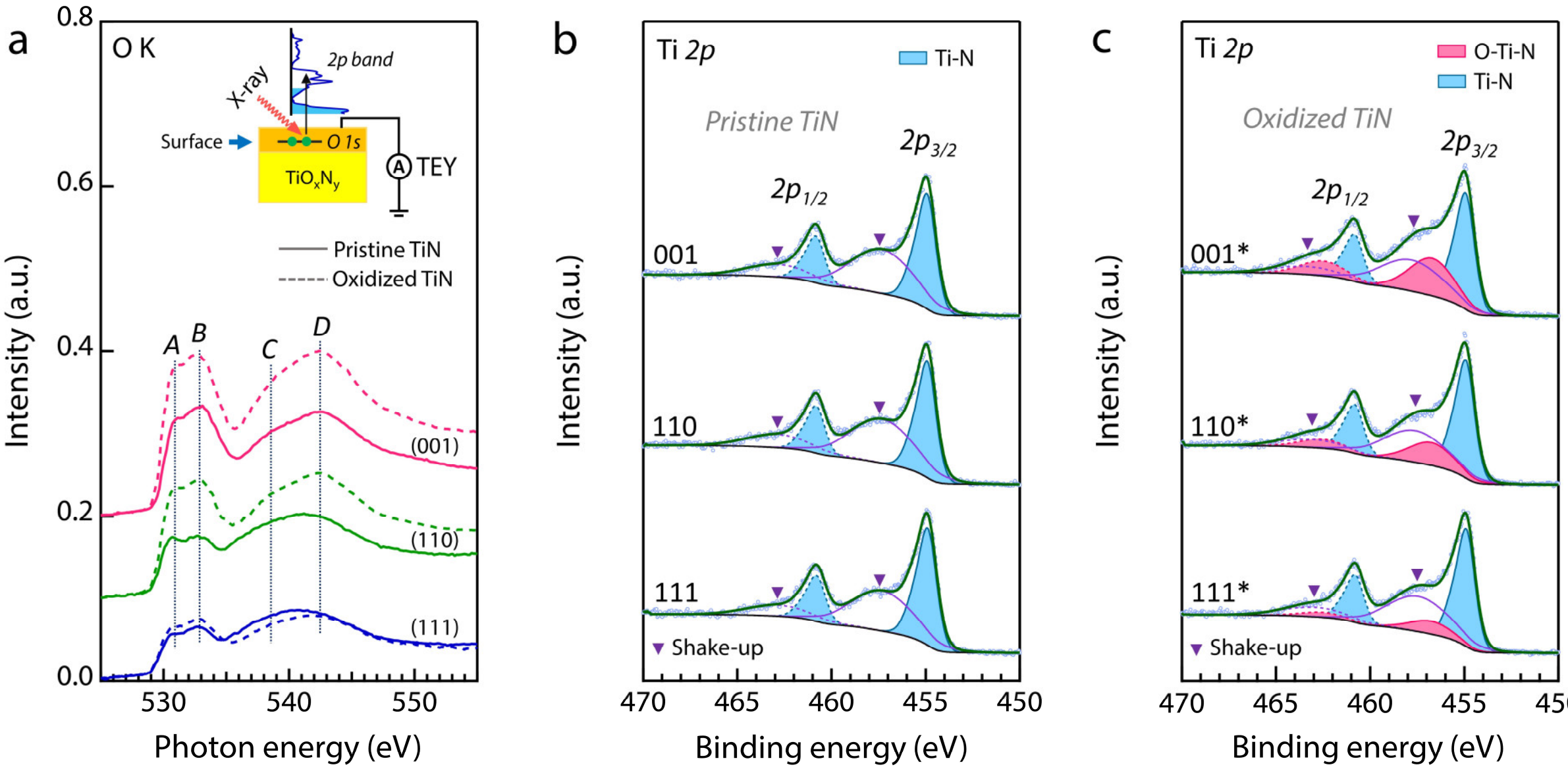

**Figure 4.** (a) O *K*-edge XAS spectra at room temperature utilizing surface-sensitive TEY detection mode. *A, B, C, D* label the dominative features of XAS spectra. (b,c) Ti 2*p* XPS spectra of pristine (b) and oxidized (c) TiN films at room temperature after removing the oxidized surfaces by etching.

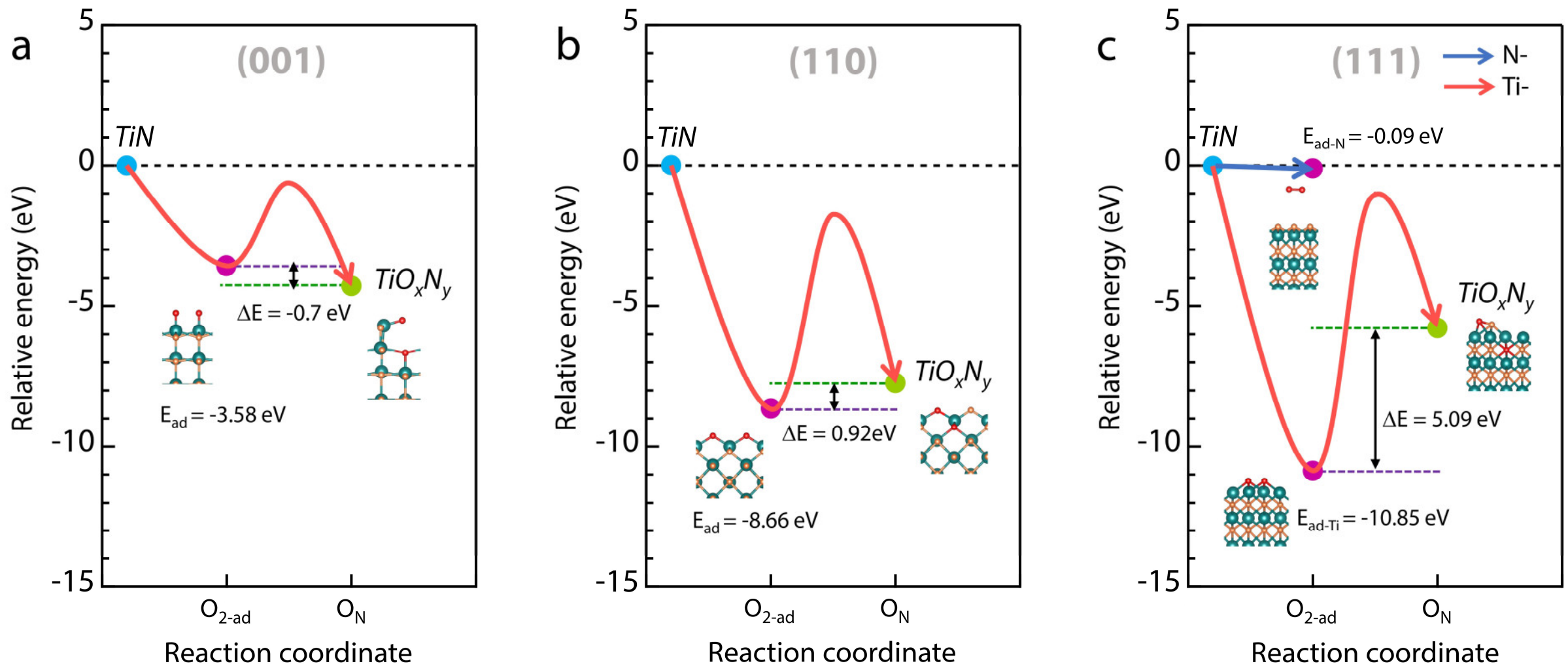

**Figure 5.** The adsorption energy ($E_{ad}$) of an $O_2$ molecule and the energy cost ($\Delta E$) for the O-N exchange on the (a) (001)-, (b) (110)-, and (c) (111)-oriented surfaces.

For Table of Contents Use Only

# Crystal orientation-dependent oxidation of epitaxial TiN films with tunable plasmonics

Ruyi Zhang,[†,‡] Qian-Ying Ma,[†,§] Haigang Liu,[∥] Tian-Yu Sun,[†] Jiachang Bi,[†] Yang Song,[†,‡] Shaoqin Peng,[†,‡] Lingyan Liang,[†] Junhua Gao, [*,†] Hongtao Cao,[†] Liang-Feng Huang,[*,†] and Yanwei Cao[*,†,‡]

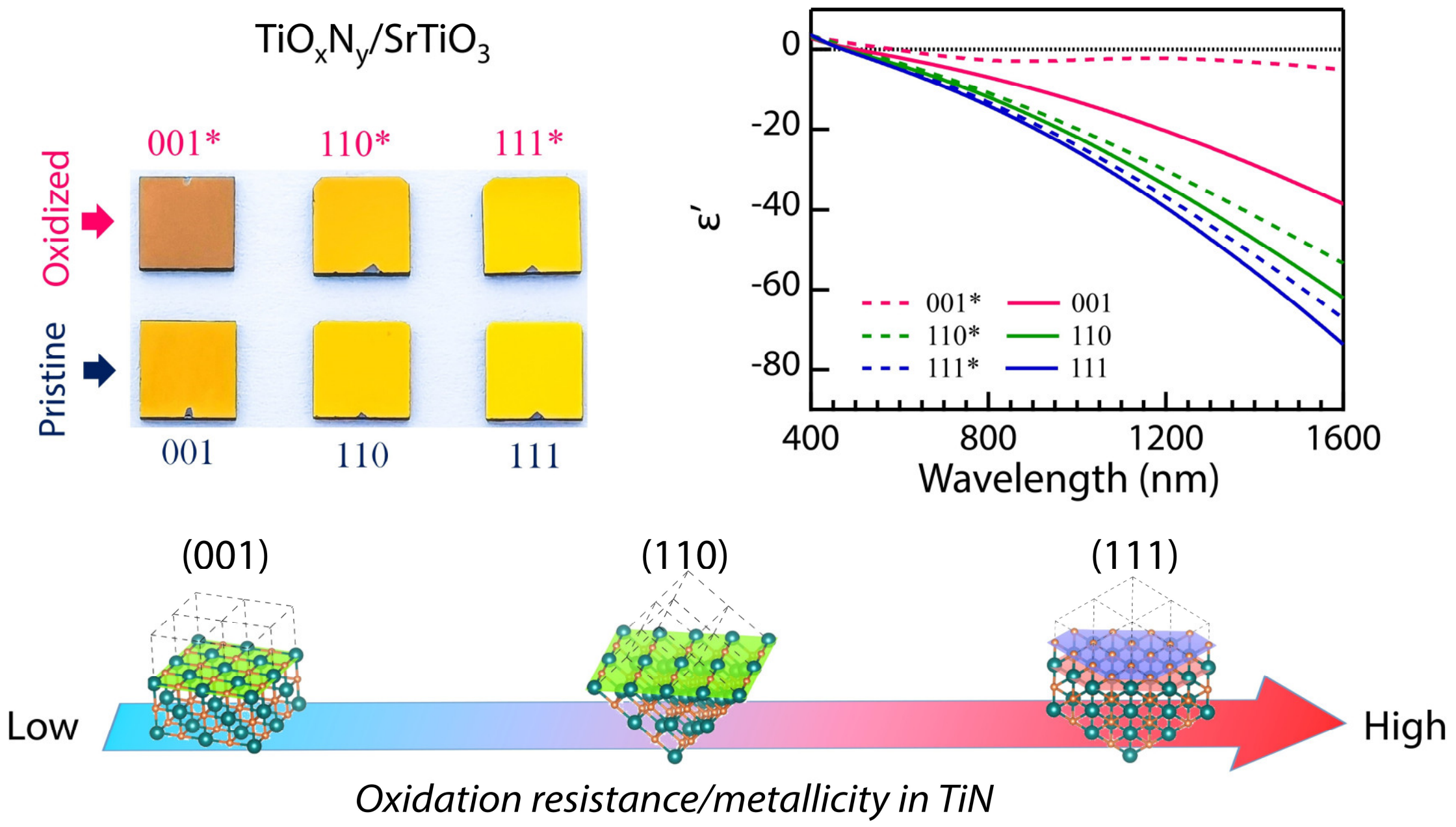

Both plasmonic performance and oxidation resistance of epitaxial TiN films follow the order of (001) < (110) < (111).